# Spectral Statistics in the Lowest Landau Band

Mario Feingold, Yshai Avishai

*Dept. of Physics, Ben-Gurion University, Beer-Sheva 84105, Israel*

and Richard Berkovits

*Dept. of Physics, Bar-Ilan University, Ramat Gan 52900, Israel*

**Abstract**

We study the spectral statistics in the center of the lowest Landau band of a 2D disordered system with smooth potential and strong transverse magnetic field. Due to the finite size of the system, the energy range in which there are extended states is finite as well. The behavior in this range can be viewed as the analogue of the Anderson metal-insulator transition for the case of the Hall system. Accordingly, we verify recent predictions regarding the exponent of the asymptotic power law of $\Sigma^2(\bar{N})$, $\gamma$, and that of the stretched exponential dominating the large $s$ behavior of the spacings distribution, $\alpha$. Both the relations, $\alpha = 1 - \gamma$, and $\gamma = 1 - \frac{1}{\nu d}$ where $\nu$ is the critical exponent of the localization length and $d$ is the dimension, are found to hold within the accuracy of our computations. However, we find that none of several possible models of the entire spacings distribution correctly describes our situation. Finally, for very large $\bar{N}$, $\bar{N} > 60$, we find a new regime in which $\Sigma^2(\bar{N})$ behaves as a power law with an unexpectedly large power, $\gamma_1 = 1.38 \pm 0.02$.

PACS numbers: 71.30.+h, 73.40.Hm, 05.40.+j

## 1. Introduction

The microscopic description of disordered systems often relies on various types of Random Matrices. These are in general different from the traditional Gaussian Random Matrix Ensembles, GOE, GUE and GSE, which were introduced in the context of nuclear physics in the early fifties. The Gaussian Ensembles are mainly used to describe the statistical properties of the spectra of strongly interacting, complex quantum systems such as nuclei for example. Consequently, such ensembles are assumed to be invariant under similarity transformations and therefore, their members are basically structureless matrices with elements, $h_{ij}$, that are independent, Gaussian distributed random variables. On the other hand, in disordered systems, the central phenomenon is the localization of eigenvectors in configuration space. While such phenomenon is in general absent in the Gaussian Ensembles, it is characteristic of random matrices with off-diagonal elements that decay as a function of $|i - j|$. Such matrices have a preferred representation and are often referred to as *banded*.

In the definition of the Gaussian Ensembles all the structure of the physical system is ignored, except for its symmetry with respect to time reversal. Accordingly, their prediction power is based on the assumption of universality, although the extent of the corresponding universality class is only qualitatively determined. For example, disordered systems in the metallic regime have been shown to share most of the spectral properties of the Gaussian Ensembles for small enough energy intervals. In particular, the spacings distribution, $P(s)$, is of Wigner type and the number variance, $\Sigma^2(\bar{N})$, is logarithmic. Disordered systems in the insulating regime however, belong to a different universality class. Namely, their $P(s)$ is Poisson and $\Sigma^2(\bar{N}) = \bar{N}$ for $\bar{N} < N_T$. In order to understand the spectral characteristics of the insulating regime, one can think of the sample as being composed of subsystems of the size of the localization length, $\xi$. While the spectrum of each subsystem is of metallic type, the full spectrum consists of a *random* superposition of many such spectra and therefore is entirely uncorrelated, or in other words, Poissonian. The universal behavior in both the metallic and the insulat-ing regimes is restricted to a certain, system dependent, energy range known as the Thouless energy, $E_T = \hbar/\tau$, where $\tau$ is the time it takes an electron to diffuse through the sample.

It was recently suggested that a third type of spectral behavior which is neither Wigner nor Poisson should be expected in the neighborhood of a metal-insulator (MI) transition.[1,2] Using a perturbative approach, it was shown that for a $d$-dimensional system, inside an energy interval, $\delta E$, which is centered on the transition energy, $E_c$, and in which $\xi > L$, where $L$ is the system size,

$$P(s) \propto \exp(-A_d \beta s^{2-\gamma}) \quad \text{for} \quad s \to \infty, \quad (1)$$

and

$$\Sigma^2(\bar{N}) \propto \frac{1}{\beta} \bar{N}^\gamma \quad \text{for} \quad \bar{N} \to \infty. \quad (2)$$

If $\nu$ is the critical exponent of $\xi$, $\xi \propto (E - E_c)^{-\nu}$, then $\gamma = 1 - \frac{1}{\nu d}$. Finally, $\beta = 1, 2, 4$ according to the symmetry with respect to time reversal.

Several numerical studies of the spectral statistics in the neighborhood of the MI transition in the 3D Anderson model were also performed. In particular, in Ref. 3 the family of $P(s)$ functions that interpolate between the Wigner and the Poisson form is obtained for a sample of finite size. All the curves in this family cross in one point at $s \simeq 2$ and their maxima appear to be equally high.[4] Moreover, $P(s) = c_1 s$ for $s \ll 1$ and for the critical distribution ($W = 16$), $c_1 \simeq 2.04$. Scaling theory implies that the family of $P(s)$ is parametrized by a single variable, $P(s, L/\xi_\infty)$, where $\xi_\infty = \lim_{L \to \infty} \xi_L$. In the thermodynamic limit, $L \to \infty$, only the Wigner, the Poisson and the critical $P(s)$ can occur. As the size of the system grows, the other curves gradually migrate towards the Wigner distribution for $W < W_c$ and toward the Poisson $P(s)$ when $W > W_c$. In another study, Evangelou[5] has shown that

$$P_V(s) = c_1 s^\beta \exp(-c_2 s^{1+\alpha}), \quad (3)$$

with $\beta = 1$ gives the best fit to the numerically obtained critical $P(s)$ for $c_1 = 2.65$, $c_2 = 1.47$ and $\alpha = 0.31$. Moreover, the fitted curve appears to be in good agreement with the numerical one. For the value of $\alpha$ the error was estimated around $\pm 0.06$. Consequently, it was suggested that for $\nu = 1$ this confirms the prediction of Eq. (1), $\alpha = 0.33$. Although the value of $\nu$ is not yet fully agreed upon, most studies indicate that $\nu \simeq 1.35$. This, in turn, leads to $\alpha \simeq 0.25$ which still agrees with the numerical value. When instead a power law was fitted to $\ln P(s)$ for $2 \le s \le 4$, the resulting $\alpha$ was quite low, that is, $\alpha \simeq 0.12$. The behavior of the critical $\Sigma^2(\bar{N})$ is also shown in Ref. 5. It indeed displays power law behavior

with $\gamma \simeq 0.88$. It therefore appears that there is better support for the relation

$$\gamma = 1 - \alpha, \quad (4)$$

where the large $s$ value of $\alpha$ is considered than for the relation between $\gamma$ and $\nu$. As was pointed out in Ref. 6, Eq. (4) holds quite generally. In a large enough energy interval, $s$, the number of levels, $N$, can be assumed to be Gaussian distributed. If the corresponding average is $\bar{N}$ and the variance is $\Sigma^2$ then the probability of having no levels in $s$, $P(s)$, is

$$P(s) \propto \exp(-\bar{N}^2/\Sigma^2) = \exp(-cs^{2-\gamma}), \quad (5)$$

where Eq. (2) was used for $\Sigma^2$ and $s$ is in units of the mean spacing.[7] Therefore, Eq. (4) also holds for separate ranges of $\bar{N}$ and the corresponding ranges of $s$. Another feature of $\Sigma^2(\bar{N})$, shown in Fig. 2(b) of Ref. 5 but otherwise ignored, is that a power law behavior is observed also for very small $\bar{N}$ although with a slightly smaller exponent, $\gamma_0 \simeq 0.71$. The change in the exponent occurs at $\bar{N} = \bar{N}_0 \approx 2.5$. Moreover, it is interesting to notice that $\gamma_0$ and the value of $\alpha$ obtained from fitting Eq. (3) to the data satisfy Eq. (4) well within the corresponding errors. As we shall see in the next Section, such fit is dominated by the $s < 2$ range and accordingly, the resulting $\alpha$ is determined by the same range of $\bar{N}$ as $\gamma_0$. On the other hand, it is surprising that the Gaussian approximation should hold here. Another fit of the critical $P(s)$ with Eq. (3) was done in Ref. 8. Since here the analytical forms of $c_1$ and $c_2$ obtained from the normalization and $\bar{s} = 1$ requirements were used, only a one parameter fit was performed. The resulting exponent was $\alpha = 0.20 \pm 0.03$ which corresponds to a slightly too large value of $\nu$, $\nu = 1.7 \pm 0.2$. It is not clear whether the difference between the value of $\alpha$ obtained in Ref. 5 and that of Ref. 9 is due to the quality of the fit or the different parameters of the model.

In this paper we study the level statistics in the neighborhood of another MI transition, namely, that occurring in two dimensional disordered systems in the presence of a strong transverse magnetic field. Such systems have been found to display quantized Hall conductance which is a consequence of the divergence of $\xi$ at the center of the disorder broadened Landau level. There is considerable evidence that the corresponding critical exponent is $\nu = \frac{7}{3}$ and that this value is more reliable than the one for the Anderson transition. Therefore, the Hall system is preferable for verifying the predictions of Eqs. (1 - 2). On the other hand, since here extended states can only be found precisely at the center of the band, it is not clear whether these predictions apply. However, in a system of finite size the critical region is broadened into a range of energy, $\Delta E$, where $\xi > L$. Inside $\Delta E$ the Hall system is critical in a way similar to that of the Anderson type system inside $\delta E$.

Numerical studies of the corresponding spectral statistics of Hall systems[9] were done prior to the work of Refs. 1 and 2. It was found that the spacings distribution varies from an almost GUE $P(s)$ in the center of the Landau band to a Poisson one at its edges. The corresponding transition was also observed in the behavior of the spectral rigidity, $\Delta_3(\bar{N})$. On the quantitative side, it was suggested that the ensemble of Pandey and Mehta[12,13] that interpolates between GOE and GUE can reproduce the $P(s)$ obtained close to the center of the band. The members of this ensemble, $H_{PM}$, are of the form

$$H_{PM}(\alpha_{PM}) = H_S + i\alpha_{PM} H_A, \quad (6)$$

where $H_S$ belongs to the GOE matrix, $H_A$ is a real antisymmetric random matrix and $0 \leq \alpha_{PM} \leq 1$ such that, $H_{PM}(1) = $ GUE. Presently, only the $P(s)$ for the case of $2 \times 2$ matrices is known

$$P_{PM}(s) = \frac{s}{4v^2\sqrt{1-\alpha_{PM}^2}} \exp\left(-\frac{s^2}{8v^2}\right)$$
$$\times \text{erf}\left(\sqrt{\frac{1-\alpha_{PM}^2}{8\alpha_{PM}^2 v^2}} s\right), \quad (7)$$

where $v$ is determined by the requirement that $\bar{s} = 1$,

$$v = \sqrt{\frac{\pi}{8}} \left(\alpha_{PM} + \frac{1}{\sqrt{1-\alpha_{PM}^2}} \arctan \frac{\sqrt{1-\alpha_{PM}^2}}{\alpha_{PM}}\right)^{-1}. \quad (8)$$

In particular, for the $P(s)$ calculated from the first 20 levels around the center of the band (out of a total of 200 levels in each sample and with 700 samples being used) good agreement was obtained with $P_{PM}(\alpha_{PM} = 0.515)$. Notice however that for large $s$ the error function in $P_{PM}$ approaches unity and thus $\gamma = 0$, in contradiction with the prediction of Refs. 1 - 2. Qualitative studies of the $P(s)$ for Hall systems were also reported in Refs. 10 and 11.

## 2. Results

In order to obtain the spectrum of an electron in a 2D random potential and a strong magnetic field, $\mathbf{B} = B\hat{z}$, we numerically diagonalize the Hamiltonian matrix, $H_{kk'}$, in a lowest Landau level basis, $\Phi_k(x,y)$. In the Landau gauge,

$$H = \left(-i\frac{d}{dx} - \frac{y}{l_B^2}\right)^2 - \frac{d^2}{dy^2} + V(x,y), \quad (9)$$

where $l_B = \sqrt{\hbar c/eB}$ is the magnetic length and

$$V(x,y) = \sum_{n,m} v_{nm} \exp\left(\frac{(x-x_n)^2 + (y-y_m)^2}{2\sigma^2}\right). \quad (10)$$

The $v_{nm}$ coefficients are uncorrelated random numbers chosen from a uniform distribution with zero mean and variance $w^2/12$ and $(x_n, y_m) = (-a/2 + (n-1)a, -a/2 + (m-1)a)$ where $a$ is the lattice constant. Moreover, the system is restricted to a square of size $L$ such that $N_l \equiv L/a$ is an integer and accordingly, $1 \leq n, m \leq N_l$. The corresponding Lowest Landau Level (LLL) states are $K$-fold degenerate in the absence of disorder, $K = L^2/(2\pi l_B^2)$, and are chosen to satisfy periodic boundary conditions[14]

$$\Phi_k(x,y) = (Ll_B)^{-1/2} \pi^{-1/4} \sum_{j=-\infty}^{\infty} e^{[i(2\pi k/L + jL/l_B^2)x]}$$
$$\times e^{-[(y-jL)/l_B - 2\pi k l_B/L]^2/2}, \quad (11)$$

where $k$ is integer and $-K/2 \leq k \leq K/2$. Finally, shifting the origin of the energy such that the LLL is at $E = 0$, one obtains

$$H_{kk'} = \frac{1}{L\sqrt{\pi(l_B^2 + 2\sigma^2)}} \sum_{n,m} v_{nm} \sum_{j,j'} e^{-S_{k+jK,k'+j'K}^{nm}}, \quad (12)$$

where

$$S_{kk'}^{nm} = \frac{1}{4}(l_B^2 + 2\sigma^2)(q_k - q_k')^2 + i(q_k - q_k')x_n$$
$$- \frac{1}{l_B^2 + 2\sigma^2}\left(y_m - \frac{q_k + q_k'}{2}l_B^2\right)^2, \quad (13)$$

and $q_k = \frac{2\pi}{L}k$. Since the terms in the second sum of Eq. (12) decay rapidly with $j$ and $j'$, it is sufficient to truncate the sum such that $-1 \leq j, j' \leq 1$. The parameters determining the strength and shape of the disorder are $w$, $a$ and $\sigma$. We first require that the overlap between two adjacent impurities be $\exp(-\frac{1}{2})$. Specifically, consider the overlap integral

$$R(a,\sigma) = \int dx dy \exp\left(-\frac{x^2 + y^2}{2\sigma^2}\right)$$
$$\times \exp\left(-\frac{(x-a)^2 + y^2}{2\sigma^2}\right). \quad (14)$$

Then

$$R(a,\sigma)/R(0,\sigma) = \exp\left(-\frac{a^2}{4\sigma^2}\right) = \exp\left(-\frac{1}{2}\right), \quad (15)$$

which leads to

$$\sigma^2 = \frac{a^2}{2} . \qquad (16)$$

Moreover, we assume $a = l_B$ such that there is one impurity within an area of $l_B^2$. Let us now fix the strength of the disorder, $w$. Notice that after projecting the Hamiltonian on the lowest Landau level, the value of $w$ only determines the energy scale of the problem. For convenience we let the width of the density of states, $\mu_2$, be unity

$$\mu_2(a^2 = 2\sigma) = \frac{w^2}{24\pi a^2(l_H^2 + a^2)} = 1 . \qquad (17)$$

Finally, the value of $B$ in our computations corresponds to $K = 1019$ and we have used 160 different samples. In order to avoid the tails of the Landau band, we restrict ourselves to the energy range $|E| < E_m = 2.5$, which in turn we divide into $N_E = 5$ equal intervals. In the third interval, $|E| < 0.5$, both the density of states and the localization length are almost constant. Therefore, in what follows, we shall regard the properties of the spectrum in this interval as critical. In order to compute the spacings distribution, we have first unfolded the spectrum normalizing each individual spacing to the average of the $N_F = 5$ adjacent ones. In Fig. 1 we show the resulting $P(s)$ for the case where the $N_B = 119$ spacings closest to $E = 0$ were used. Three different theoretical curves are fitted to the numerical experiment: (a) the distribution of Eq. 3 with $\beta = 2$, $P_V(s)$, (b) the Pandey-Mehta distribution of Eqs. (7 - 8), $P_{PM}(s)$, and (c) the Robnik distribution[15], $P_R(s)$.

The Robnik distribution interpolates between the Poisson $P(s)$ and the GUE one using a $2 \times 2$ random matrix, $H_R(\alpha_R)$. For $\alpha_R = 0$, the matrix is diagonal and $P(s) = e^{-s}$, while for $\alpha_R \to \infty$ it approaches the GUE. In practice, when $\alpha_R \approx 1$ the corresponding spacings distribution is very close to that of GUE. Specifically,

$$P_R(s) = P_+(s) + P_-(s), \qquad (18)$$

where

$$P_\pm(s) = \frac{s}{4a^2} \frac{e^{\lambda^2}}{\lambda^2} \int_0^{s/a} x \frac{e^{\mp((S/a)^2 - x^2)^{1/2}}}{((S/a)^2 - x^2)^{1/2}} e^{-x^2/4\lambda^2}$$
$$\times \left[1 \pm \operatorname{erf}\left(\frac{1}{2\lambda}((S/a)^2 - x^2)^{1/2} \mp \lambda\right)\right] dx, \quad (19)$$

$\lambda = \alpha_R/a$ and $a$ is determined by the $\bar{s} = 1$ constraint.

Fitting $P_V(s)$ to the numerically obtained $P(s)$ with the assumption that the number of spacings in each of the 38 bins is a Gaussian distributed random variable leads to $\alpha = 0.26 \pm 0.02$ and $\chi^2 = 85.0$ (see Fig. 1(a)).

For the case of a least squares fit with 37 degrees of freedom this corresponds to a negligible confidence limit, $CL = 0.7 \cdot 10^{-3}\%$. One might wonder whether the large value of $\chi^2$ obtained is a consequence of ignoring additional sources of error besides those of statistical origin. The most likely such source is the unfolding procedure. In order to verify this possibility we modify the extent of the smoothing interval, $N_F$, to 7. Comparing the resulting $P(s)$, $P_7(s)$, against the original one, $P_5(s)$, and using twice the statistical variance we obtain $\chi^2 = 29.1$ for which $CL \simeq 62\%$. Therefore, despite significantly larger statistics than usually encountered in Quantum Chaos where in general only a single sample is studied, here as well, the accuracy of the unfolding procedure appears to be practically irrelevant. One concludes that $P_V(s)$ is inappropriate for the description of our numerical results. Nevertheless, it is interesting to notice that most of the contribution to the $\chi^2$ comes from the small $s$ range. In particular, we find that the partial $\chi^2$ from the first 10 bins, $s < 1$, is 48.2 while that from the last 18 bins, $s > 2$, is only 18.0. This fact suggests that the large $s$ tail of $P_V(s)$ could be in agreement with $P(s)$ despite the fact that the two distributions are different from each other. We therefore separately fit the $s > 2$ tail of $P_V(s)$ and this leads to $\alpha = 0.30 \pm 0.02$, $\chi^2 = 13.9$ and $CL \simeq 74\%$. This result is statistically reliable and, at the same time, is in contradiction with the prediction of Ref. 2 for $\nu = 7/3$ which is $\alpha = 0.214$. Alternatively, using this result to determine the value of $\nu$ one obtains, $\nu = 1.7 \pm 0.1$.

While $P_V(s)$ does incorporate the expected behavior of the true $P(s)$ at both small and large $s$, it is largely an interpolation formula for $s = O(1)$. In other words, there is no physical argument to support such behavior as that of $P_V(s)$ for intermediate values of $s$. On the other hand, the Pandey-Mehta distribution interpolates between the GOE $P(s)$ and the GUE one along the most natural path in the space of $2 \times 2$ random matrices. At first, it would appear that there should be no GOE component in the description of a Hall system. However, such component is needed in order to incorporate those states that are extended, $\xi > L$, but are relatively insensitive to changes of the flux and accordingly, behave almost as if time-reversal invariant. We therefore fit $P_{PM}(s)$ to the numerical $P(s)$ (see Fig. 1(b)) and obtain $\alpha_{PM} = 0.18 \pm 0.01$ and $\chi^2 = 94.4$. In other words, the Pandey-Mehta distribution is even less appropriate that $P_V(s)$ to describe the level spacings at the center of the Landau band. Moreover, we find that the contribution to the large value of $\chi^2$ comes equally from the entire range of $s$. Specifically, the partial $\chi^2$ for small $s$, $s < 1$ (the first 10 bins), is 23.3, while that from large $s$, $s > 2$ (last 18 bins) is 54.0. Accordingly, when the fit is restricted to the large $s$ tail, $s > 2$, we obtain $\alpha_{PM} = 0$ and $\chi^2 = 37.2$. Not only does this correspond to a confidence limit of only 0.32%, but rather than a true minimum of $\chi^2$ it is just the edge of the allowed interval of the fitting parameter, $\alpha_{PM}$. One is lead to conclude that the tail of the distribution is not of Gaussian type, that is, $\alpha \neq 1$.

If instead of computing the spacing distribution in the central (third) energy interval, we use the fifth, edge interval, a $P(s)$ that is quite close to a Poisson distribution is obtained (see Fig. 2). This is to be expected considering that the corresponding eingenstates are localized, $\xi \ll L$. On the other hand, from the knowledge of the finite size localization length, $\xi_L$, alone, there is no way one can determine the boundaries of the critical energy range. It is therefore conceivable that even in the central interval some fraction of the states are localized. Since the Pandey-Mehta distribution ignores the possibility of a Poisson component due to localized states, it is worthwhile to try to identify the presence of such component by fitting a Poisson - GUE interpolating distribution, e.g. $P_R(s)$ (see Fig. 1(c)). We obtain that $\alpha_R = 0.66 \pm 0.02$ and $\chi^2 = 604.4$ which corresponds to a negligible confidence limit. Although here, like for $P_V(s)$, most of the contribution to $\chi^2$ comes from the small $s$ range ($\chi^2(s < 1) = 365.2$, $\chi^2(s > 2) = 106.8$), the fit of the $s > 2$ tail alone leads to $\alpha_R = 0.75 \pm 0.02$, $\chi^2 = 92.8$ and $CL < 10^{-3}$%. This clearly excludes the Robnik distribution as a candidate for the description of our results. Moreover, it indicates that in order to derive an appropriate distribution one has to use the fact that some of the eingenvectors are localized. It is likely that such distribution is similar to the one encountered in a 1D disordered system when $\xi \approx L$ (see Refs. 16 - 20). The failure of $P_R(s)$ to incorporate localization is further stressed by fitting it to the edge $P(s)$ of Fig. 2. The best fit is obtained for $\alpha_R = 0.75 \cdot 10^{-3}$ with $\chi^2 = 85.9$ and $CL < 10^{-3}$%.

As already mentioned, we do not have a clear estimate of the critical energy range. It is therefore instructive to verify the behavior of the tail of $P(s)$ as we shrink the energy interval around the center of the band. We therefore progressively reduce the number of energy levels around $E = 0$, $N_B$, and fit $P_V(s)$ for $s > 2$. The resulting values of $\alpha$ are given in Tbl. 1. One would expect that as we lower $N_B$ and correspondingly, the statistics becomes worse, the value of $\chi^2$ would go down and the error would increase. We see however that while the error does indeed increase, the $\chi^2$ goes first way up before returning to about the same value at $N_B = 9$ as at $N_B = 119$. Therefore, although in Tbl. 1 the value of $\alpha$ is varying beyond what our error estimates would allow, it is reasonable to expect that the values at $N_B = 9$ and $N_B = 119$ are the most reliable. Using the former together with the prediction of Ref. 2 to obtain the critical exponent, leads to $\nu = 1.7 \pm 0.4$, which still disagrees with the corresponding theoretical value, $\nu = 7/3$. However, we incline to view this disagreement as being a consequence of our inaccurate numerical procedure rather than indicating the failure of the theory of Ref. 2.

We now turn to the study of $\Sigma^2(\bar{N})$. In Fig. 3 we show the result obtained for the third, central interval. Ignoring the range, $0.8 < \bar{N} < 2.3$ in which some oscillations occur, one can distinguish three different power law ranges: I. $0.07 < \bar{N} < 0.8$, II. $2.3 < \bar{N} < 60$, and III. $60 < \bar{N} < 350$. The best fitting power law curves give $\gamma_0 = 0.74 \pm 0.01$, $\gamma = 0.77 \pm 0.02$ and $\gamma_1 = 1.38 \pm 0.02$, respectively. Moreover, the corresponding natural logarithms of the coefficients are $C_0 = -0.69 \pm 0.02$, $C = -1.05 \pm 0.04$, and $C_1 = -3.5 \pm 0.1$, respectively. As in Ref. 5, $\gamma_0$ and the value of $\alpha$ obtained from the fit of the entire $P_V$ satisfy Eq. (4). This relation also holds for $\gamma$ and the $\alpha$ of the $s > 2$ tail of $P_V$ but only if the large error bar of the $N_B = 9$ case is used. On the other hand, the relation between $\gamma$ and $\nu$ is in good agreement with the prediction of Ref. 1. The value of $\gamma_1$ however, bears practically no relation to any previous result or theoretical prediction. While presently the origin of this regime is not clear, we have observed similar behavior in other models of the Hall system as well. Finally, in Fig. 4 we show the $\Sigma^2(\bar{N})$ obtained from the last, band edge energy interval. Here, one has to modify the $\bar{N}$ such as to account for the finite size of the sample, that is, $\bar{N} \rightarrow \bar{N}_1 \equiv \bar{N}(1 - \bar{N}/K)$. Assuming as before a power law dependence for $\Sigma^2(\bar{N}_1)$, the best fit is obtained for a power of $\gamma = 0.97 \pm 0.005$ and a coefficient whose natural logarithm is $C = 0.007 \pm 0.02$. This is extremely close to the prediction of the Poisson ensemble but not identical, indicating that even in the tails of the Landau band there are a few states with relatively large $\xi$.

The finite size correction used for the data of Fig. 4 assumes that the levels are uncorrelated. In fact, the effect of this correction is not too large. Specifically, the best power law fit to $\Sigma^2(\bar{N})$ is $\gamma = 0.953 \pm 0.005$ and $C = 0.03 \pm 0.02$. This however, is because in Fig. 4, $\bar{N}$ only goes up to about 60. On the other hand, the analogous finite size correction might significantly modify Fig. 3 where $\bar{N}$ reaches almost 350. In particular, it could modify the value of $\gamma_1$. Since it is known that levels in this range are strongly correlated, it is not clear what the proper finite size correction should be. It would be interesting though to study the dependence of $\gamma_1$ on the system size.

## 3. Conclusions

In summary, we have shown that the predictions of Refs. 1 - 2 hold for the spectral statistics at the center of the lowest Landau band to the accuracy of our numerics. Moreover, it was pointed out that aside from the small and large $s$ behavior of $P(s)$, there is no real understanding of the spacings distribution interpolating between the metallic and the insulating regimes. In fact, the three most natural theoretical distributions, $P_V(s)$, $P_{PM}(s)$ and $P_R(s)$, were shown to disagree with the numerical $P(s)$. It should be instructive to further compare with the distribution derived by Haake and Lenz[21] from the way in which the Dyson gas relaxes to equilibrium from the configuration corresponding to a Poisson ensemble. However, in all these distributions one makes no explicit use of the fact that the eigenvectors of the system are localized. The simplest way to incorporate the presence of localization is in the spirit of renormalization theory. Namely, one assumes that the system is composed of independent blocks of size $\xi$ for which the corresponding spectrum has a Wigner $P(s)$. The spectrum of the entire system is then a random superposition of the block spectra and the $P(s)$ of such superpositions can be derived.[22] Unfortunately, it turns out that the resulting $P(s)$ has a nonvanishing value at $s = 0$ in contradiction with the observed form. Recently, a new ensemble which has a preferred basis and thus, is reminiscent of the situation occurring in the case of localization, has been shown to be analytically tractable.[23] However, the $P(s)$ for this ensemble is not yet known.

Finally, we have shown that for very large $\bar{N}$, $\Sigma^2(\bar{N})$ grows with an unexpected power that is significantly larger than unity.

We would like to thank I. Dana, D. Eliyahu, B. Horovitz and Y. Meir for useful discussions. This work was supported by the Israel Science Foundation administered by the Israel Academy of Sciences and Humanities.

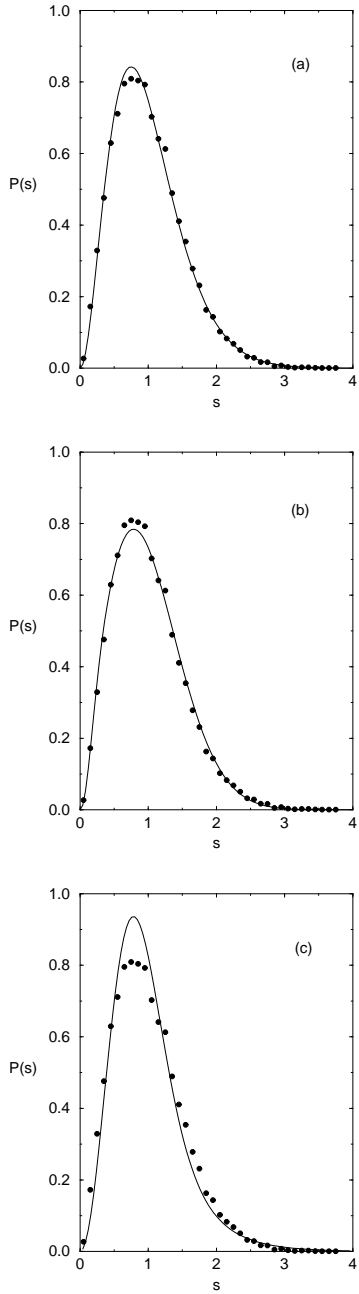

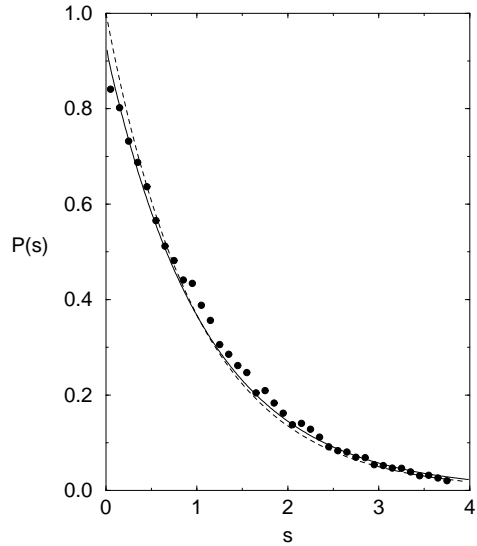

Fig. 2. The spacings distribution for the $N_B = 119$ levels closest to $E = 2$ which, in turn, is close to the edge of the band (•). The best fitting Robnik distribution (solid) and the Poisson distribution (dashed) are also shown.

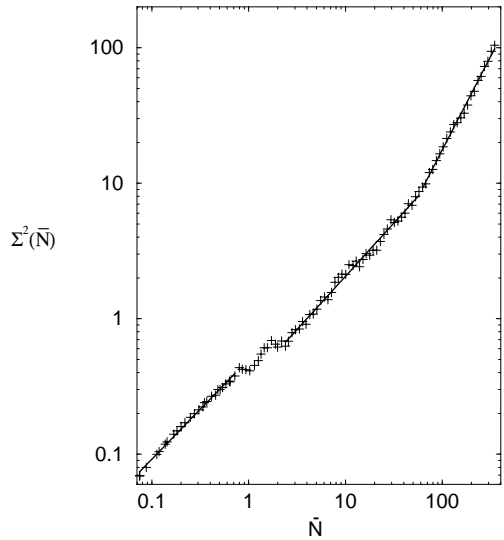

Fig. 1. The spacings distribution for the $N_B = 119$ levels closest to the center of the Landau band (•). The line is the corresponding best fit in the entire $0 < s < 4$ range of a) $P_V(s)$, b) the Pandey-Mehta distribution, and c) the Robnik distribution.

Fig. 3. The number variance, $\Sigma^2(\bar{N})$, for the central energy interval, $|E| < 0.5$ (+). The solid lines are the corresponding power law fits in the three different $\bar{N}$ ranges (see text).

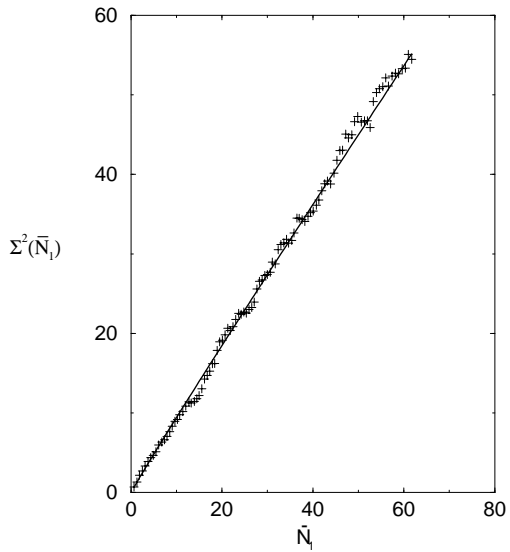

Fig. 4. Same as in Fig. 3 for the energy interval, $1.5 < E < 2.5$. Here however, the finite size corrected average number of levels, $\bar{N}_1$, is used as variable, instead of $\bar{N}$.

Table 1. The values of $\alpha$ and the corresponding $\chi^2$ for gradually decreasing energy intervals around the center of the Landau band in which the $P(s)$ was computed.

| $N_B$ | $\alpha$ | $\chi^2$ |
|---|---|---|
| 119 | $0.30 \pm 0.02$ | 13.9 |
| 59 | $0.40 \pm 0.03$ | 19.5 |
| 29 | $0.43 \pm 0.05$ | 21.2 |
| 19 | $0.37 \pm 0.06$ | 17.3 |
| 9 | $0.30 \pm 0.08$ | 13.3 |